\documentclass[preprintnumbers,amsmath,11pt,amssymb,floatfix,superscriptaddress,nofootinbib]{article}
\usepackage{array}
\usepackage{relsize}
\usepackage{cite}
\usepackage{amsmath}
\usepackage{amssymb}
\usepackage{graphicx}
\def\mytitle#1{\setcounter{equation}{0}
\setcounter{footnote}{0}
\begin{flushleft}\Large\textbf{#1}\end{flushleft}
\vspace{0.25cm}}
\def\myname#1{\leftline{{\large #1}}\vspace{-0.13cm}}
\def\myplace#1#2{\small\begin{flushleft}\textit{#1}\\
\texttt{#2}\end{flushleft}}

\def\myclassification#1{\small\noindent
{\bf PACS:}
       #1\vspace{0.5cm}}
\usepackage{graphicx}
\begin{document}

\mytitle{Violation of Universal Lower Bound for the Shear Viscosity to Entropy Density Ratio in Dark Energy Dominated Accretion }

\myname{Sandip Dutta \footnote{duttasandip.mathematics@gmail.com} and Ritabrata Biswas\footnote{biswas.ritabrata@gmail.com}}
\myplace{Department of Mathematics, The University of Burdwan, Golapbag Academic Complex, Burdwan -713104, Dist.:- Purba Barddhaman, State:- West Bengal, India.}{} 
 
\begin{abstract}
The universal lower bound of the ratio of shear viscosity to entropy density is suggested by the string theory and gauge duality for any matter. We examined the ratio of shear viscosity to entropy density for viscous accretion flow towards a central gravitating object in the presence of dark energy. The ratio appears close to the universal lower bound for certain optically thin, hot accretion flows as they are embedded by strong magnetic field. Dark energy is a kind of exotic matter which has negative pressure. So dark energy creates repulsive force between the accreting particles, which indicates that shear viscosity of the flow becomes very low. Dark energy as accreting fluid has very high entropy density. The ratio should reach near to the lowest value for dark energy accretion. We wish to study what happens to the shear viscosity to entropy density ratio for viscous dark energy accretion flow.
	

\myclassification{97.10.Gz;	51.20.+d; 51.30.+i; 95.36.+x; 97.60.Lf}
\end{abstract}
Strongly interacting quantum field theories depict a description of dual holographic natures of many states among which the black Holes (BHs) in Anti de-Sitter(AdS) space is a well known one. For this category of BHs a universal lower bound \cite{kovtun2005, policastro2001, kovtun2003, buchel2004, son2007a} is prescribed as,
\begin{equation}
\frac{\eta}{s} \geq \frac{1}{4 \pi} \frac{\hbar}{k_B},
\end{equation}
which is known as the Kovtun-Starinets-Son(KSS) bound. All known fluids for which $\frac{\eta}{s}$ is measured (even for quark-gluon plasma created at the Relativistic Heavy-Ion Collider(RHIC)) satisfy the lower bound \cite{Alberte, teaney2003, romatschke2007, song2008, romatschke2008, dusling2008, grozdanov2015, haehl2015}. This KSS lower bound is found to get related to the  minimum entropy production of a BH. Shear viscosity is popularly obtained with a retarded Green's function through the Kubo formula, which takes the form,
\begin{equation}
\eta= \lim_{\omega \to 0} \frac{1}{2\omega} \int   \exp \left\lbrace i \omega t\right\rbrace dt d\vec{x} \langle \left[ T_{xy}(t, \vec{x}), T_{xy}(o, \hat{o}) \right] \rangle,
\end{equation}
where $T_{xy}$ is the $xy$-th component of the stress tensor. This KSS lower bound is saturated by Einstein's gravity theory \cite{brigante2008a} and this enforces Einstein gravity to violate the bound on a generic small correction. Causality will also be violated. Charge dependence and causality dependence of $\frac{\eta}{s}$ in different modified gravity have been studied in different references \cite{ge2008, amoretti2015}. References \cite{ge2008, ge2009a, ge2009b} depicted the upper bound on the Gauss-Bonnet coupling constant to satisfy the theoretical lower bound of $\frac{\eta}{s}$. At low temparature, if we consider four dimensional bulk space-times (where the mass fluctuations of the metric components $\delta g_{xy}$ does not vanish), the ratio $\frac{\eta}{s}$ would tend to a constant value \cite{burikham}. It was concluded that per ``Planckian time" \cite{hartnoll, zaanen2004}, the ratio of shear viscosity to entropy density is equal to the logarithmic increase of the entropy production. If there exists a scale $\Delta$ in the zero temparature infrared theory, it is reasonable to consider the presence of a diffrent temparature independent source, $\delta g^0_{xy}=t \Delta$ that bounds  the entropy production, given by
$$ \frac{\eta}{S}\gtrsim \left( \frac{T}{\Delta}\right)^2,~~~~ \frac{T}{\Delta}\longrightarrow 0~~.$$
This new bound is violated in the reference \cite{ling}. Earlier of this decade, Jackovac \cite{jackovac2010} has examined the ratio  $\frac{\eta}{s}$ mathematically assuming some physical conditions for the spectral functions, and keeping the entropy density constant. He has found that for some systems the lower bound is not universal. Candidates of such system are quasiparticle systems with small wave function renormalization constant, high temperature strongly interacting systems, or low temperature systems with zero mass excitations.

In this letter, we wish to consider the phenomenon of viscous dark energy (DE hereafter) accretion onto a supermassive BH and construct the dependency of $\frac{\eta}{s}$ on different accretion parameters. To build up this problem, we will firstly require a brief knowledge on what is BH accretion and what DE is.

If we wish to set the simplest example, BH is actually a general relativistic space consisting of at least two singularities: one at $r=0$ and another at $r=\frac{2GM}{c^2}$ ($M$ is mass of the central object and c= velocity of light in vacuum). The space-time wrapped by the latter singularity is known as a BH space-time. Accretion (generally known as Roche lobe outflow) is a flow of diffused material or plasma towards some central gravitating object, mainly taken to be a massive star or a compact object. Following the Rayleigh's criterion, the infalling object forms a structure of diffused material in spiral motion and the viscous friction between the adjacent layers along with the gravitational energy loss heats up the material and electromagnetic radiation is thrown out. Accretion around a BH or a neutron star emits in X-ray whereas protoplanetary accretion may radiate in infrared. Accretion around a Schwarzschild type of singularity was studied by Michel \cite{Michel} where he has taken the integrations of continuity and energy flux equations and combined to form a radial inward velocity gradient (ordinary differential equation consisting $\frac{dv}{dr}$). To make the flow continuous, he has drawn the idea of critical point where both the numerator and the denominator of the equation $\frac{dv}{dr}=\frac{N}{D}$ will vanish. Only solution that passes through a critical point corresponds to material falling into (or flowing out of). Next memorable landmark in the history of accretion studies was done by Shakura and Sunyaev \cite{ShakuraSunyaev1973}. They have proposed that if there exists transport of angualar momentum, it will give birth of magnetic field, turbulence and molecular and radiative viscosity which will grow tangential stresses between adjacent layers of the flow. There will exist plasma instabilities, reconnection of magnetic field lines in regions of opposite polarity field etc, which will let the energy of the magnetic field in the disk not to exceed the thermal energy of the matter, i.e., for magnetic energy $\frac{H^2}{8 \pi}< \frac{\rho c_s^2}{2}$ and  $w_{r \phi} \sim - \rho v_s^2 \left(\frac{H^2}{4 \pi \rho c_s^2} \right)$. One among the efficiency of two of the most important mechanisms, i.e., angular momentum transport connected to another, i.e., the magnetic field and turbulence as 
$$w_{r \phi} \sim \rho c_s^2 \left(\frac{v_z}{c_s} \right) + \rho c_s^2 \left(\frac{H^2}{4 \pi \rho c_s^2}\right) = \alpha_{ss} \rho c_s^2$$
The nonlinearity which arises from the general relativity can be removed by considering Pseudo Newtonian potential. Such a potential depending on distance from BH and rotational parameter of BH was calculated in the reference \cite{pseudonewtonian} as 
\begin{equation}
F_g\left(x\right)= \frac{\left(x^2-2j\sqrt{x}+ j^2\right)^2}{x^3\lbrace \sqrt{x} \left(x-2\right)+j\rbrace ^2}~~,
\end{equation}
where $x=\frac{r}{r_g}$, $j$ is dimensionless specific angular velocity for rotating BH. We will construct our model with this particular potential.
We will consider the equation of continuity 
\begin{equation}
\frac{\partial \rho}{\partial t}+ \vec{\nabla}.\vec{v}=0
\end{equation}
and the steady state part of it for cydindrical coordinate becomes
\begin{equation}\label{equation of continuity}
\frac{d}{dx} \left(x u \Sigma\right) = 0~~.
\end{equation}
where $\Sigma$ is vertically integrated density given by,
$
\Sigma = I_c \rho_e h\left(x\right)~~,
$
when $I_c$= constant (related to equation of state of the accreting fluid)= $1$, 
$\rho_e$=density at equitorial plane, 
$h\left(x\right)$= half thickness of the disc. $u=u_x=\frac{v_x}{c}$, $v_x$ is the radially inward speed of accretion. 
Next we will consider Navier Stokes's equation for steady state given as 
\begin{equation}\label{navier_stoke}
\rho \left(\vec{u}. \vec{\Delta} \right) \vec{u}= - \vec{V} p + \rho \gamma \Delta^2 \vec{u} -\vec{F}_{Gx} ~~.
\end{equation}
Where $\vec{u}$ is the velocity vector, $p$ is the pressure, $\gamma$ is the viscosity coefficient. The radial component of this equation turns out to be 
\begin{equation}\label{radial momentum balance}
u\frac{du}{dx} + \frac{1}{\rho} \frac{dp}{dx}-\frac{\lambda^2}{x^3}+F_g\left( x\right)=0~~,
\end{equation}
where all the variables are expressed in dimensionless units $u=u_r=\frac{v}{c}$, $x=\frac{r}{r_g}$, $r_g=\frac{GM}{c^2}$, where $r$, $v$ and $\lambda$ are radial coordinate and radial velocity and angular momentum of the disc, $p$ and $\rho$ are dimensionless isotropic pressure and density.
The azimuthal momentum balance equation is given by
\begin{equation}\label{azimuthal momentum balance}
u \frac{d\lambda}{dx} = \frac{1}{x \Sigma} \frac{d}{dx} \left[ x^2 \alpha_{ss} \left(p+\rho u^2\right) h\left(x\right) \right]~~.
\end{equation}
Assuming the vertical equilibrium from the vertical component of Navier Stokes' equation we get the expression for $h(x)$ as
\begin{equation}\label{halfthikness}
h\left(x\right) = c_s \sqrt{\frac{x}{F_g}}~~.
\end{equation}
Along with all these equations we will require another equation which will depict the relation between pressure and density of the accreting fluid. For this we choose a present time observational data supported scenario of our universe. Since twenty years of now, from the distant type Ia supernova observations we got to know about the late time cosmic accelaration \cite{Riess}. An unknown form of energy, the name of which popularly coined as DE is hypothesized to permeate all over the universe homogeneously which is responsible for the accelaration of expansion of the universe. If we consider that the standard model of cosmology is correctly constructed, the best current measurements indicate that DE contributes $68\%$ of the total energy in the present day observable universe \cite{WMAP}. The density of DE is seemed to be very low $\sim 7 \times 10^{-30}$ $gm cm^{-3}$. Modified Chaplygin gas model is one among many DE representatives. The equation of state is given by
\begin{equation}\label{EoS}
p=\alpha \rho -\frac{\beta}{\rho^n}~~.
\end{equation}
Depending on differenet values of $\alpha$ and $\beta$, this fluid can mimic radiation $\left(\frac{p}{\rho} = \frac{1}{3}\right)$, barotropic fluid $\left( \frac{p}{\rho}=const. >0 \right)$, pressureless dust $(p=0)$, quintessence $\left(-1< \frac{p}{\rho}< -\frac{1}{3} \right)$ or phantom $\left( \frac{p}{\rho}<1 \right)$.
We see the speed of sound through this particular fluid is given by 
\begin{equation}\label{Eoq2}
c_s^2 = \frac{\partial p}{\partial \rho} = \alpha + \frac{\beta n}{\rho^{n+1}}~~.
\end{equation}
Thus combining (\ref{equation of continuity}) and (\ref{Eoq2}), we obtain the pressure gradient as
\begin{equation}\label{Eoq3}
\frac{1}{\rho} \frac{dp}{dx} = - \frac{2 c_s^3}{\left(n+1\right) \left( c_s^2 - \alpha \right)} \frac{d c_s}{dx} =-\frac{1}{n+1} \frac{d}{dx} \left(c_s^2 \right) - \frac{\alpha}{n+1} \frac{d}{dx} \lbrace ln \left( c_s^2 -\alpha \right) \rbrace~~.
\end{equation}
Integrating the equation (\ref{equation of continuity}) the mass conservation equation is evolved as
\begin{equation}\label{mass conservation}
\dot{M} = \Theta \rho c_s \frac{x^\frac{3}{2}}{F_g^\frac{1}{2}} u~~,
\end{equation}
where $\Theta$ is geometrical constant.\\
Replacing the value of $\rho$ from equation (\ref{Eoq2}) in (\ref{mass conservation}) and differentiating the whole term we get a ordinary  differential equation for sound speed,
\begin{equation}\label{differential equation for c}
\frac{d c_s}{dx} = \left( \frac{3}{2x} -\frac{1}{2F_g} \frac{dF_g}{dx} + \frac{1}{u} \frac{du}{dx} \right) \left\lbrace \frac{\left( n+1 \right) c_s \left( c_s^2 -\alpha \right)}{\left( 1-n \right) c_s^2 + \alpha \left( n+1 \right)} \right\rbrace~~ .
\end{equation}
The same for the specific angular momentum is obtained from (\ref{azimuthal momentum balance}) and (\ref{EoS}) as,
\begin{multline}\label{differential equation for lambda}
\frac{d \lambda}{dx} = \frac{x \alpha_s}{u} \left[ \frac{1}{2} \left( \frac{5}{x} - \frac{1}{F_g} \frac{dF_g}{dx} \right) \left\lbrace \frac{\left(n+1 \right) \alpha - c_s^2}{n} + u^2 \right\rbrace \right.\\
\left. + 2 u \frac{du}{dx} + \left\lbrace \left( \frac{\left(n+1 \right) \alpha - c_s^2}{n} + u^2 \right)  \frac{1}{c_s} -\left( c_s^2 +u^2 \right) \left( \frac{1}{n+1} \frac{2 c_s}{c_s^2 - \alpha} \right) \right\rbrace \frac{dc_s}{dx} \right]~~. 
\end{multline}
Now (\ref{radial momentum balance}) will give us the inward radial speed gradient as,
\begin{equation}\label{differential equation for u}
\frac{du}{dx} = \frac{\frac{\lambda^2}{x^3}- F_g\left( x \right) + \left( \frac{3}{x} - \frac{1}{F_g} \frac{dF_g}{dx} \right) \frac{ c_s^4}{\lbrace \left( 1-n \right) c_s^2 + \alpha \left( n+1 \right)\rbrace}}{u - \frac{2 c_s^4}{u \lbrace \left( 1-n \right) c_s^2 + \alpha \left( n+1 \right)\rbrace}}~~.
\end{equation}

When a particle starts to move towards a BH, at infinite distance, the value of radial inward velocity is almost zero and when it reaches the BH's event horizon, the inward falling speed becomes equal to that of light. So, somewhere in beteween (say at $x=x_c$) the fluid speed becomes equal to such a value that the denominator of the equation (\ref{differential equation for u}) vanishes. We will call this point as a critical point (anologus to sonic point). Now, to make the flow continuous at $x=x_c$ the numerator should also vanish. By the virtue of these two properties, $N(x_c)=D(x_c)=0$, we can obtain the velocity gradient by using L'Hospital's rule. At $x=x_c$ we get a quadratic equation of $\frac{du}{dx}$ in the form
\begin{equation}\label{quadratic}
A \left( {\frac{du}{dx}}\right)_{x=x_c}^2 + B \left( \frac{du}{dx} \right)_{x=x_c} + C =0~~.
\end{equation}
Where \\
$A= 1 + \frac{1}{c_{sc}^2} - \frac{4 \left( n+1 \right) \left( c_{sc}^2 -\alpha \right)}{{\left\lbrace \left( 1-n \right) c_{sc}^2 +\alpha \left( n+1 \right) \right\rbrace}^2}$,\\
$B= \frac{4 \lambda u}{x_c^3} + \frac{2 u_c c_{sc} \left( 1-n \right)}{ \left\lbrace \left( 1-n \right) c_{sc}^2 +\alpha \left( n+1 \right) \right\rbrace} - \frac{2 u_c \left( n+1 \right) \left( c_{sc}^2 -\alpha \right)}{{\left\lbrace \left( 1-n \right) c_{sc}^2 +\alpha \left( n+1 \right) \right\rbrace}^2} \left\lbrace \frac{3}{x_c} - \frac{1}{F_g} \left(\frac{dF_g}{dx}\right)_{x=x_c} \right\rbrace + \\
\left[ \left( c_{sc}^2 +u_c^2 \right) \left( \frac{1}{n+1} \frac{2 c_{sc}}{c_{sc}^2 - \alpha} \right)-\left\lbrace \frac{3}{x_c} - \frac{1}{F_g} \left(\frac{dF_g}{dx}\right)_{x=x_c} \right\rbrace \frac{4 u_c }{c_{sc}{\left\lbrace \left( 1-n \right) c_{sc}^2 +\alpha \left( n+1 \right) \right\rbrace}}-  \left\lbrace \frac{\left(n+1 \right) \alpha - c_{sc}^2}{n} + u^2 \right\rbrace  \frac{1}{c_{sc}} \right] \left\lbrace \frac{ u_c \left( n+1 \right) \left( c_{sc}^2 -\alpha \right)}{2 c_{sc}^3} \right\rbrace $\\
and $C= D+ E+ F$~~.\\
The values of $D$, $E$ and $F$ are,\\
$D= \left[ \left( c_{sc}^2 +u_c^2 \right) \left( \frac{1}{n+1} \frac{2 c_{sc}}{c_{sc}^2 - \alpha} \right)-\left\lbrace \frac{3}{x_c} - \frac{1}{F_g} \left(\frac{dF_g}{dx}\right)_{x=x_c} \right\rbrace \frac{4 u_c }{c_{sc}{\left\lbrace \left( 1-n \right) c_{sc}^2 +\alpha \left( n+1 \right) \right\rbrace}}-  \left\lbrace \frac{\left(n+1 \right) \alpha - c_{sc}^2}{n} + u^2 \right\rbrace  \frac{1}{c_{sc}} \right] \\
\left\lbrace \frac{3}{2x_c} -\frac{1}{2F_g} \left(\frac{dF_g}{dx} \right)_{x=x_c} \right\rbrace \left\lbrace \frac{\left( n+1 \right) c_{sc} \left( c_{sc}^2 -\alpha \right)}{\left( 1-n \right) c_{sc}^2 + \alpha \left( n+1 \right)} \right\rbrace $~~,\\
$E= \left\lbrace \frac{3}{x_c} - \frac{1}{F_g} \left(\frac{dF_g}{dx}\right)_{x=x_c} \right\rbrace \frac{u_c c_{sc} \left( 1-n \right)}{{\left\lbrace \left( 1-n \right) c_{sc}^2 +\alpha \left( n+1 \right) \right\rbrace}} + \left(\frac{dF_g}{dx}\right)_{x=x_c} + \left\lbrace \frac{1}{F_g^2} \left(\frac{dF_g}{dx} \right)^2_{x=x_c} - \frac{1}{F_g} \left(\frac{d^2 F_g}{dx^2}\right)_{x=x_c} - \frac{3}{x_c^2}\right\rbrace \frac{u_c^2}{2}$~~,\\
$F= \frac{\lambda \alpha_{ss}}{x_c^2 u_c} \left\lbrace \frac{1}{F_g} \left(\frac{dF_g}{dx}\right)_{x=x_c} - \frac{5}{x_c} \right\rbrace \left\lbrace \frac{\left(n+1 \right) \alpha - c_{sc}^2}{n} + u_c^2 \right\rbrace  $~~.

Again at $x=x_c$, the two algebraic equations $D(x_c)=0$ and $N(x_c)=0$ will provide us the relation between $u_c$ and $c_{sc}$ and the value of $c_{sc}$ respectively (if $\lambda$ is fixed to some $\lambda_c$ at $x_c$, which will give us a physical solution), i.e., we will be able to determine $u_c= u \mid _{x=x_c}$. Using these boundary conditions we may obtain the solution of the equations (\ref{differential equation for c}), (\ref{differential equation for lambda}) and (\ref{differential equation for u}) \cite{Sandip1}.

It is followed from the reference \cite{Sandip1} that viscous Chaplygin gas accretion increases the strength of outflowing wind and decreases the effective disc length as we increase the viscosity. This says that the feeding up property of accretion disc is negatively working if DE of modified Chaplygin gas type is accreting.
From the equations (\ref{EoS}) and (\ref{equation of continuity}) we get the density equation as 
\begin{equation}\label{density}
\rho= \frac{2.285 \times 10^{-21} \times \sqrt{\frac{F_g}{x_c^3}}}{u_c c_{sc}}~~.
\end{equation}
In the reference \cite{Sandip2} the variation of density has been studied and it is followed that for DE accretion a density fall in the accretion branch is found. As we increase the viscosity, the fall occurs at further distances from the BH. So the accretion density is weakend by DE if viscosity is working as an inhibitor.

Next we will come to the central part of this letter. We can write the energy equation of the above set of equations as \cite{mukhopadhyay2013} 
\begin{equation}\label{energy_equation}
uT \frac{ds}{dx}= q_{vis}+ q_{mag}+ q_{nex}-q_{rad}~~,
\end{equation}
where $T$ is the temparature of flow, $S$ is entropy per volume. $q_{vis}, q_{mag},$ and $ q_{nex}$ respectively define the energies released per unit volume per unit time due to viscous dissipation, magnetic dissipation and thermonuclear reactions. $q_{rad}$ indicate the energy radiated away per unit volume per unit time by various cooling processes like synchrotron, bremsstrahlung and inverse comptonization of soft photons and the energy absorbed per unit volume per unit time due to thermonuclear reactions. So the entropy density of the flow can be expressed as 
\begin{equation}\label{entropy_density}
s=\int{\frac{q_{vis}+ q_{mag}+ q_{nex}-q_{rad}}{u T} dx}~~.
\end{equation}
The turbulent kinematic viscosity $\gamma$ can be scaled linerly with sound speed ($c_s$) in the flow and half thickness ($h$) of the disc, hence $\gamma=\alpha_{ss} c_s h$. Therefore the shear viscosity has the form
\begin{equation}
\eta= \alpha_{ss} \rho c_s h~~.
\end{equation}
The $\frac{\eta}{s}$ ratio for DE accretion is found to be 
\begin{equation}\label{etaovers}
\frac{\eta}{s}=\frac{\alpha_{ss} c_s^2 \sqrt{\frac{x_c}{F_g}} \left(\rho \right)^{(n+1)}}{\left(\rho \right)^{(n+1)} \left(1+ \alpha\right)- \beta}~~.
\end{equation}

\begin{figure}[h!]
\centering
~~~~~~~Fig $1.1.a$~~~~~~~\hspace{3 in}~~~~~~~~~Fig $1.1.b$
\includegraphics*[scale=0.6]{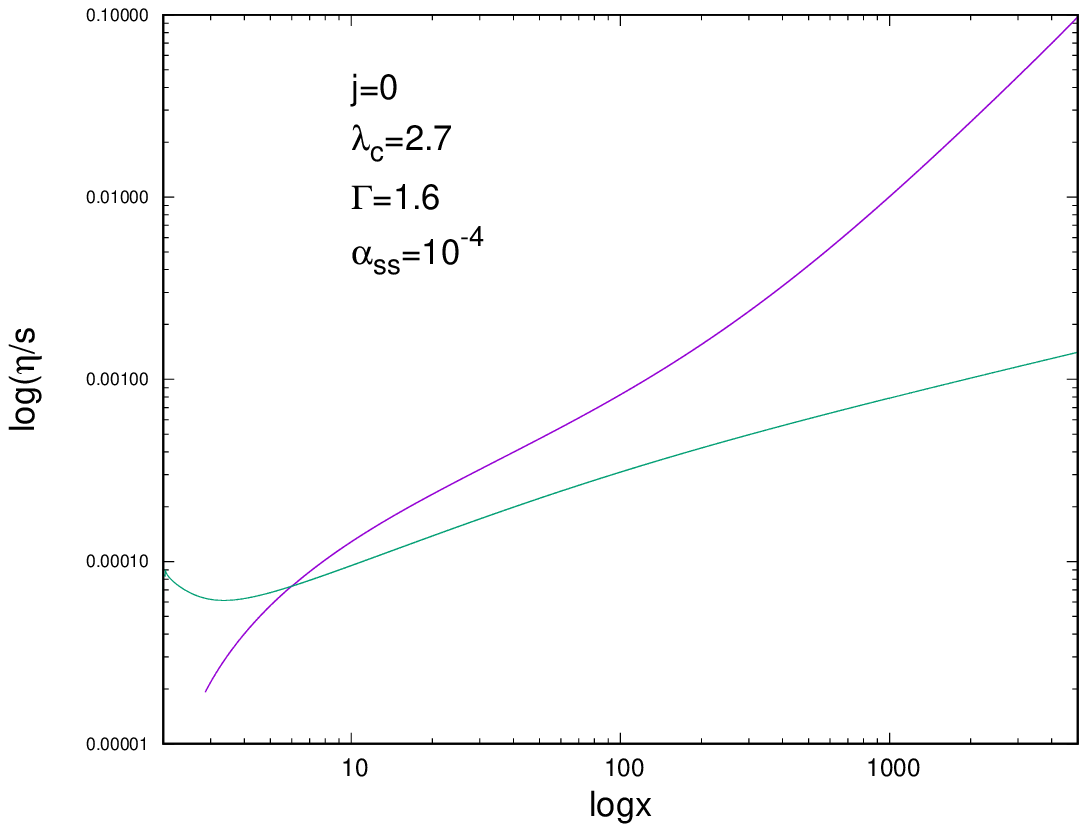}~~
\hspace*{0.1 in}
\includegraphics*[scale=0.6]{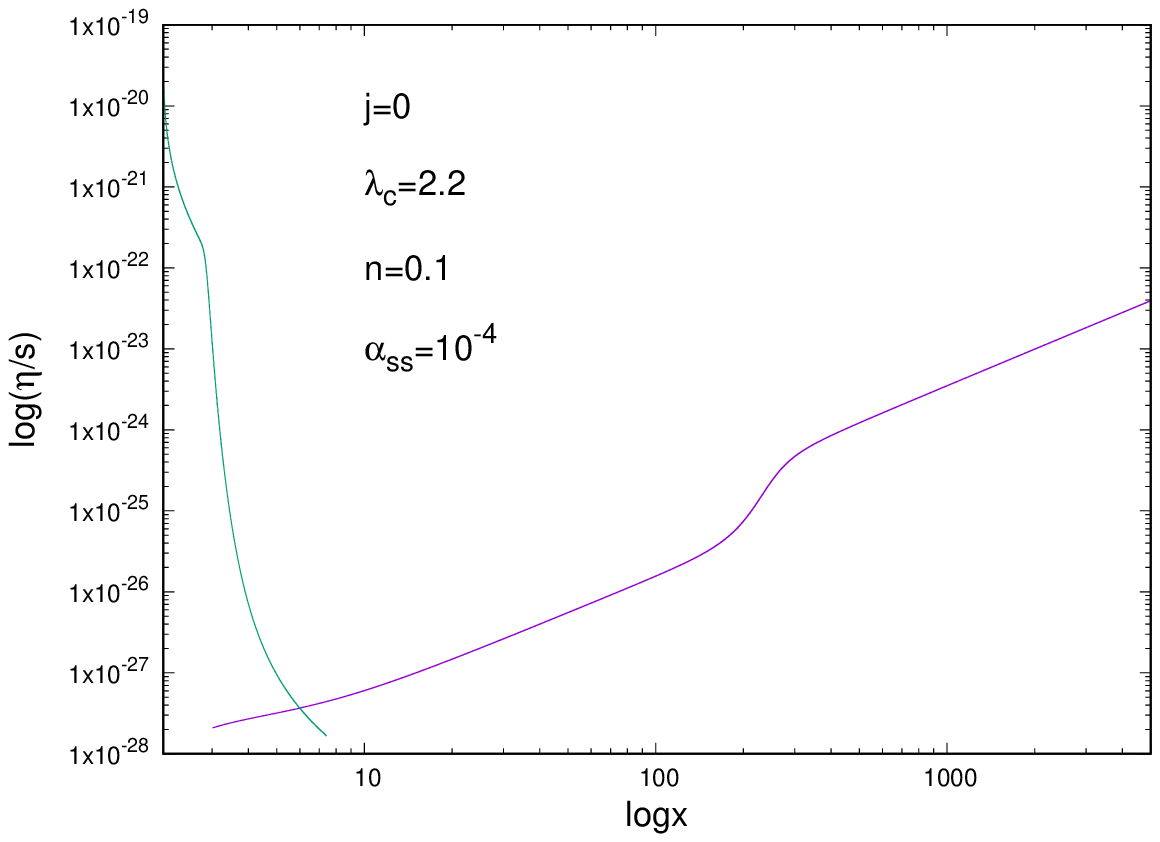}~~\\

~~~~~~~Fig $1.2.a$~~~~~~~\hspace{3 in}~~~~~~~~~Fig $1.2.b$
\includegraphics*[scale=0.6]{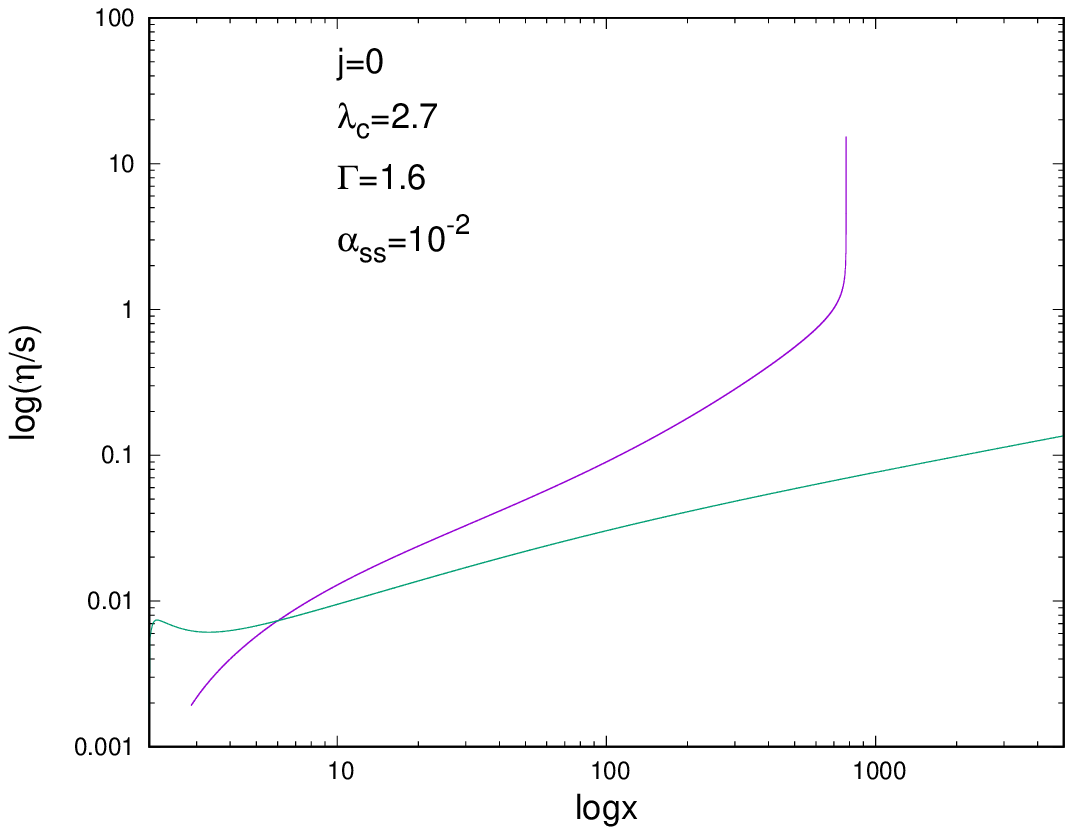}~~
\hspace*{0.1 in}
\includegraphics*[scale=0.6]{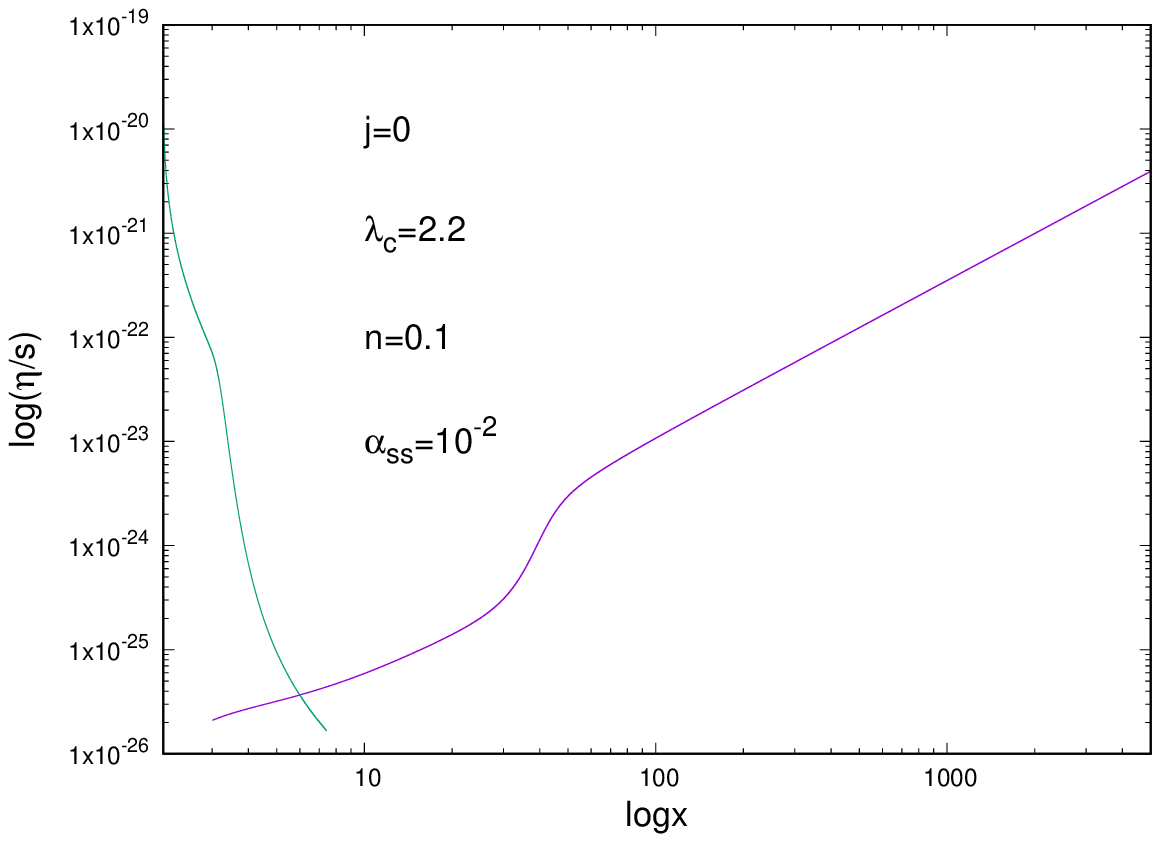}~~\\
\it{Figures $1.1.a$ and $1.2.a$ are plots of $log(\frac{\eta}{s})$ vs $log(x)$ for viscous accretion disc flow with viscosity $\alpha_{ss}=10^{-4}$ and  $\alpha_{ss}=10^{-2}$ respectively, around a non-rotating BH for adiabatic case. Figures $1.1.a$ and $1.2.a$ are plots of $log(\frac{\eta}{s})$ vs $log(x)$ for viscous accretion disc flow with viscosity $\alpha_{ss}=10^{-4}$ and  $\alpha_{ss}=10^{-2}$ respectively, around a non-rotating BH for MCG. Accretion and wind curves are depicted by purple and green lines respectively.}
\end{figure}

\begin{figure}[h!]
\centering
~~~~~~~Fig $2.1.a$~~~~~~~\hspace{3 in}~~~~~~~~~Fig $2.1.b$
\includegraphics*[scale=0.6]{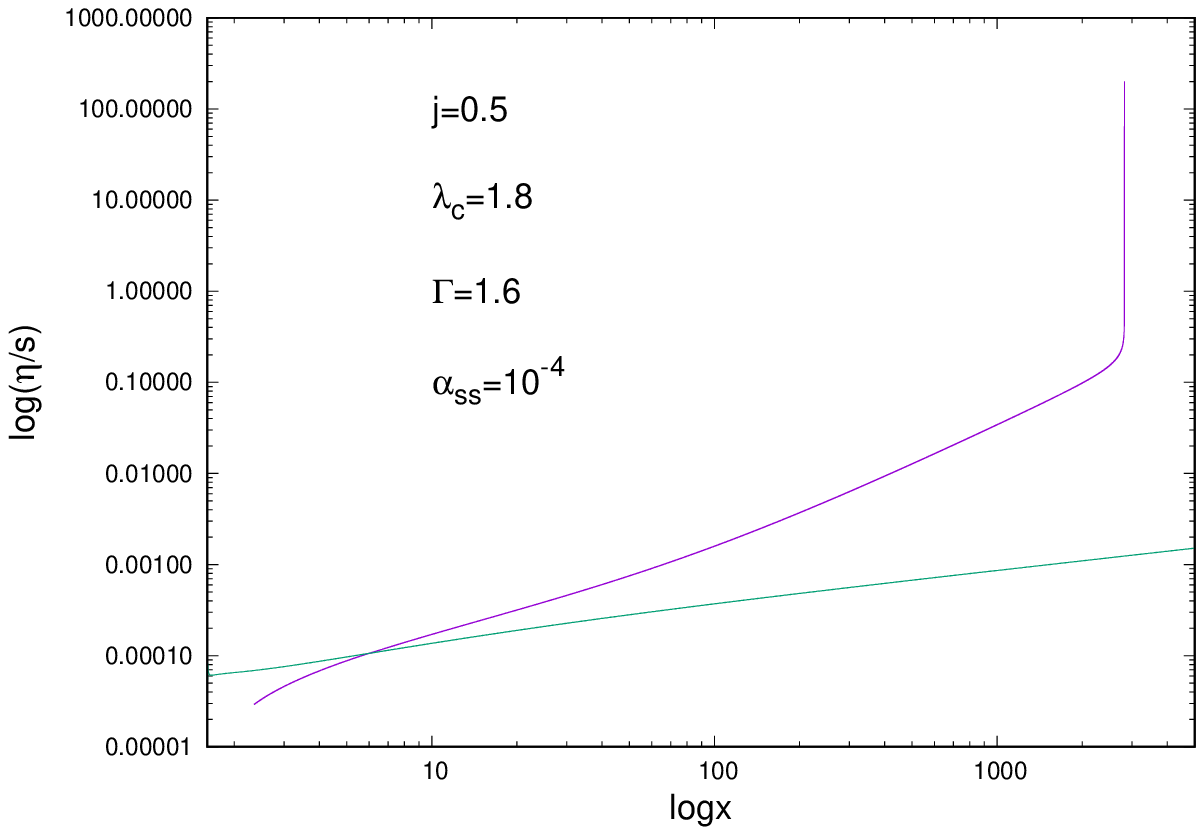}~~
\hspace*{0.1 in}
\includegraphics*[scale=0.6]{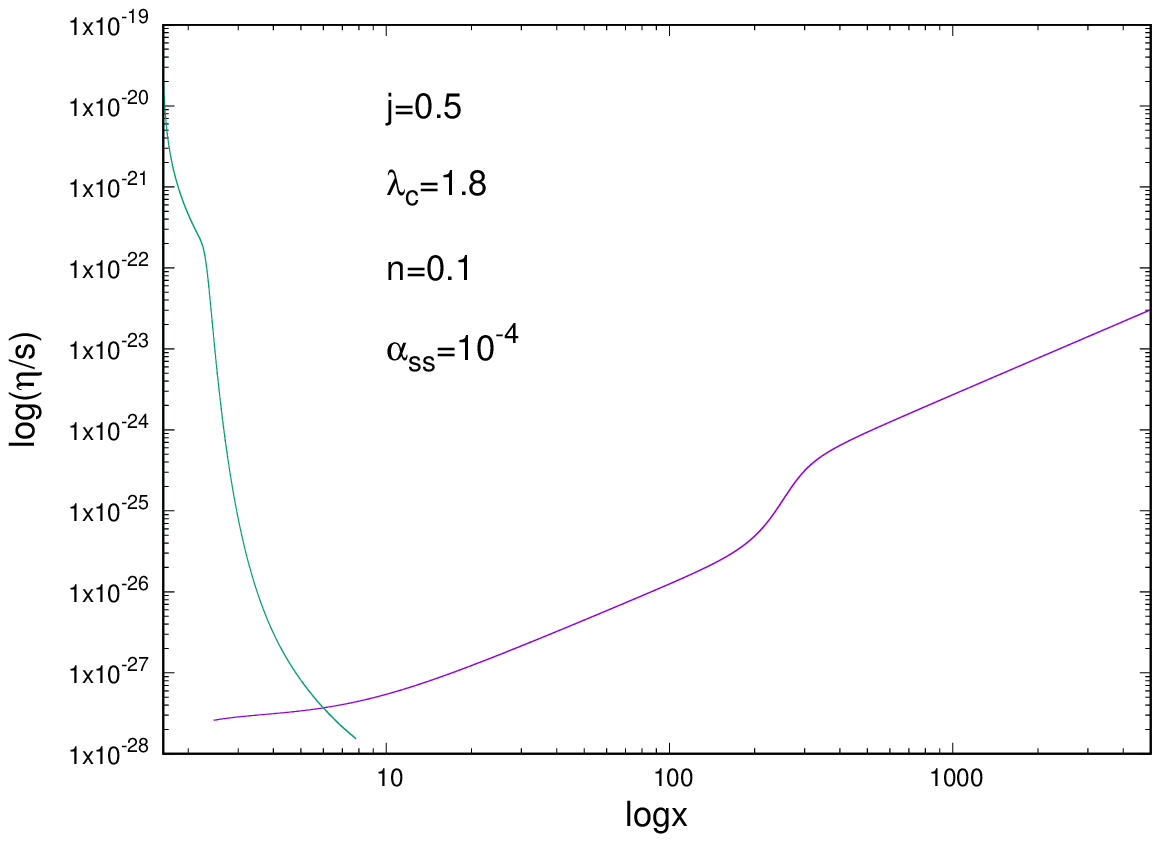}~~\\

~~~~~~~Fig $2.2.a$~~~~~~~\hspace{3 in}~~~~~~~~~Fig $2.2.b$
\includegraphics*[scale=0.6]{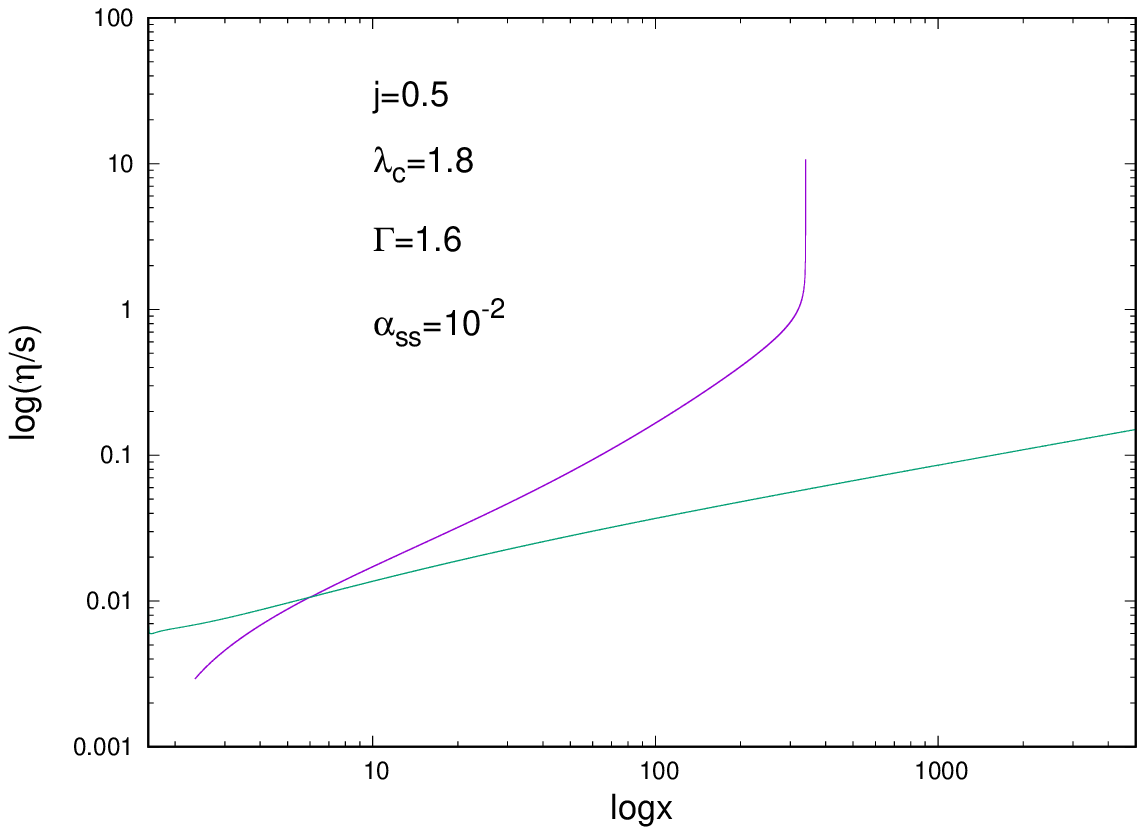}~~
\hspace*{0.1 in}
\includegraphics*[scale=0.6]{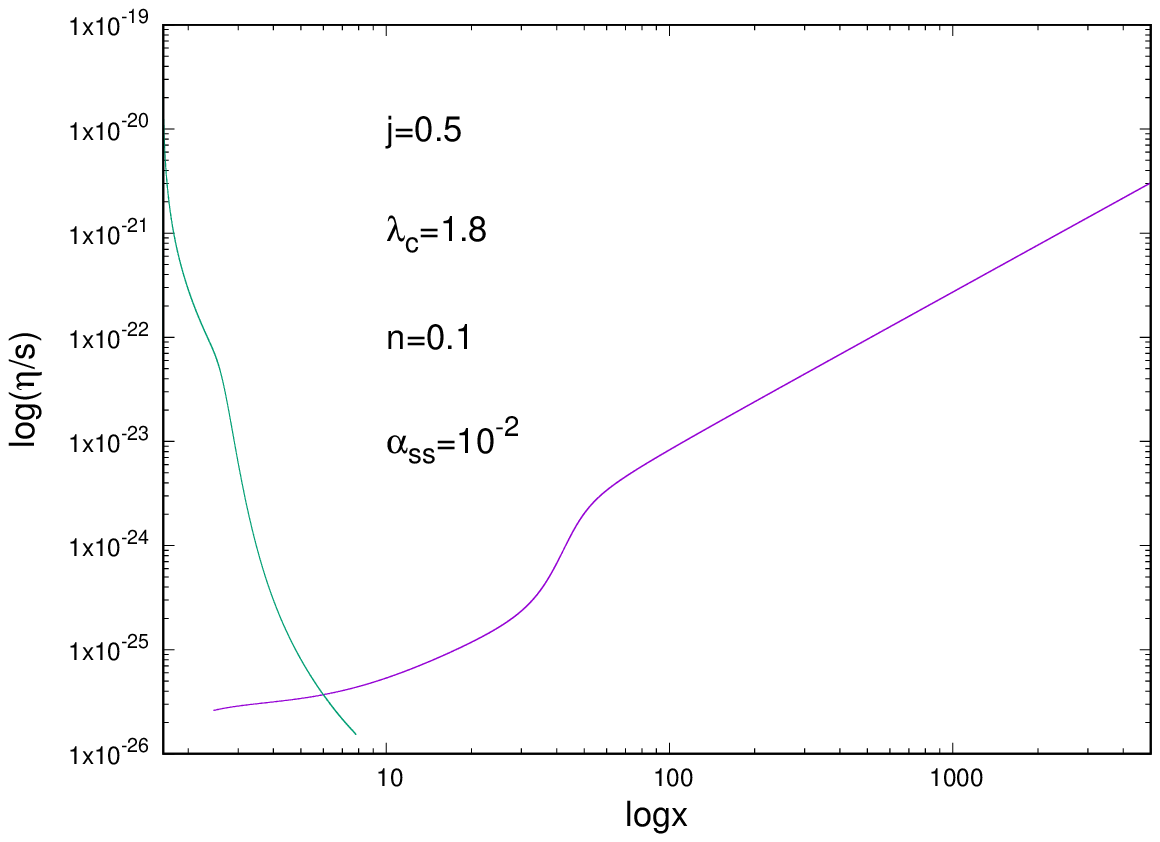}~~\\
\it{Figures $2.1.a$ and $2.2.a$ are plots of $log(\frac{\eta}{s})$ vs $log(x)$ for viscous accretion disc flow with viscosity $\alpha_{ss}=10^{-4}$ and  $\alpha_{ss}=10^{-2}$ respectively, around a rotating BH with $j=0.5$ for adiabatic case. Figures $2.1.a$ and $2.2.a$ are plots of $log(\frac{\eta}{s})$ vs $log(x)$ for viscous accretion disc flow with viscosity $\alpha_{ss}=10^{-4}$ and  $\alpha_{ss}=10^{-2}$ respectively, around a rotating BH with $j=0.5$ for MCG. Accretion and wind curves are depicted by purple and green lines respectively.}
\end{figure}
We have plotted $log \left( \frac{\eta}{S}\right)$ vs $log (x)$ in figures $1.1.a$-$3.2.b$. Accretion and wind curves are depicted by purple and green lines respectively. Figures $1.1.a$ - $1.2.b$ are for accretion (around a nonrotating BHs) of viscous fluid with $\alpha_{ss}=10^{-4}$ (Fig $1.1.a$ and $ 1.1.b$) and $\alpha_{ss}=10^{-2}$ (Fig $1.2.a$ and $1.2.b$) of adiabatic gas (Fig $1.1.a$ and $1.2.a$) and modified Chaplygin gas (Fig $1.1.a$ and $1.1.b$) respectively. For low viscosity adiabatic flow's accretion and wind both branches have decreasing $\frac{\eta}{s}$ ratio as we go towards the BH. The lowest value of $\frac{\eta}{s}$ is found to be of the order of $10^{-5}$. Wind branch's $\frac{\eta}{s}$ is higher in region $x< x_c$ and lower in the region $x>x_c$. A robust difference is observed if we consider modified Chaplygin gas as the accreting fluid. $\frac{\eta}{s}$ ratio for accretion branch is decreasing function of $(-x)$ whereas that for wind branch is a steepy increasing function of $(-x)$. Very near to the BH the ratio for accretion branch becomes of the order of $10^{-28}$ whereas far (but at infinite distance) from the BH, where the wind velocity becomes equal to that of light, the ratio is also of the order of $10^{-28}$ for the wind branch. This speculates that if exotic matter's accretion is consider we can reach very near to the prescribed lower limit of $\frac{\eta}{s}$ (still we are $100$ times higher than the limit).
\begin{figure}[h!]
	\centering
	~~~~~~~Fig $3.1.a$~~~~~~~\hspace{3 in}~~~~~~~~~Fig $3.1.b$
	\includegraphics*[scale=0.6]{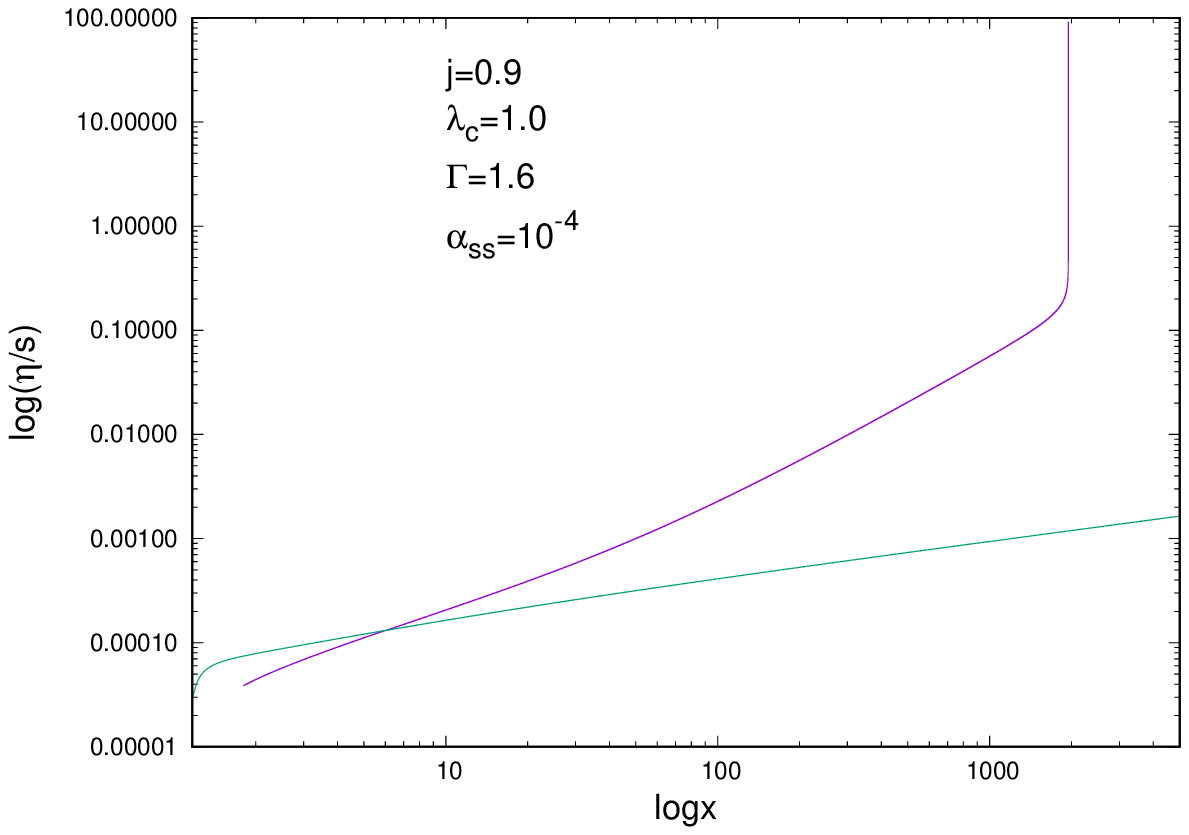}~~
	\hspace*{0.1 in}
	\includegraphics*[scale=0.6]{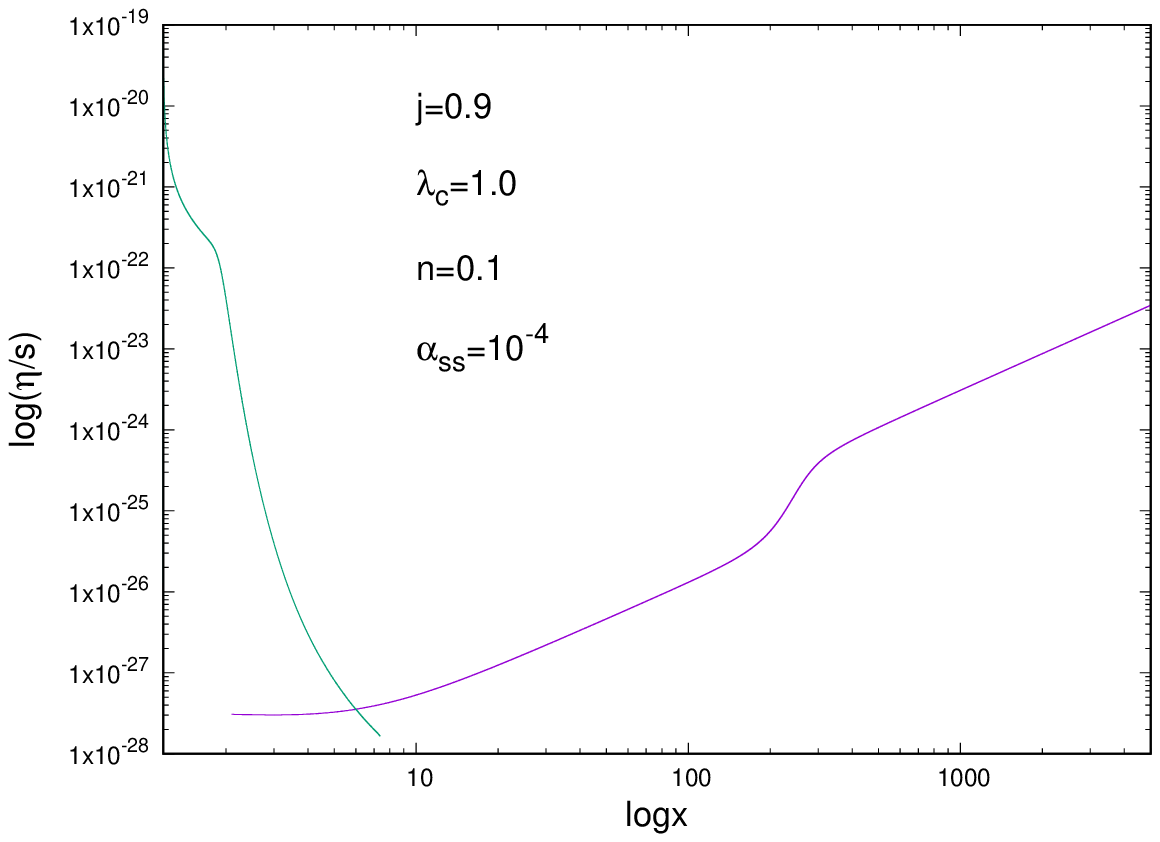}~~\\
	
	~~~~~~~Fig $3.2.a$~~~~~~~\hspace{3 in}~~~~~~~~~Fig $3.2.b$
	\includegraphics*[scale=0.6]{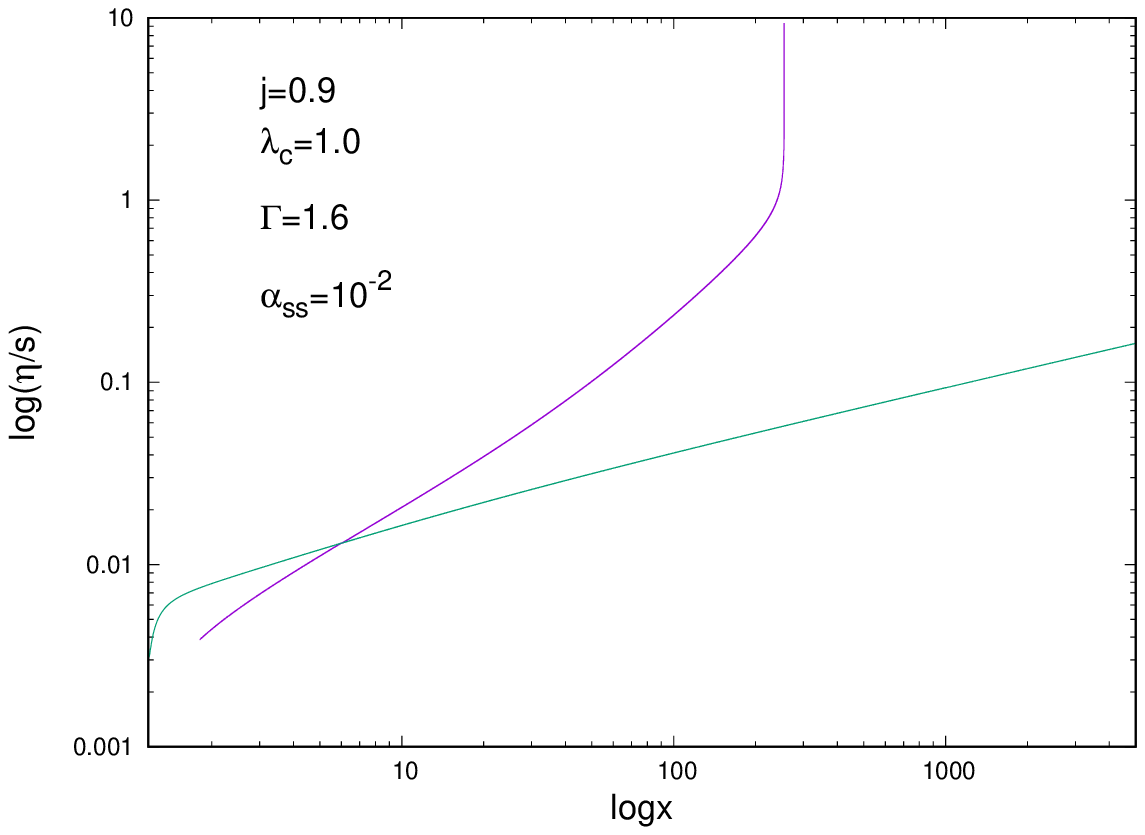}~~
	\hspace*{0.1 in}
	\includegraphics*[scale=0.6]{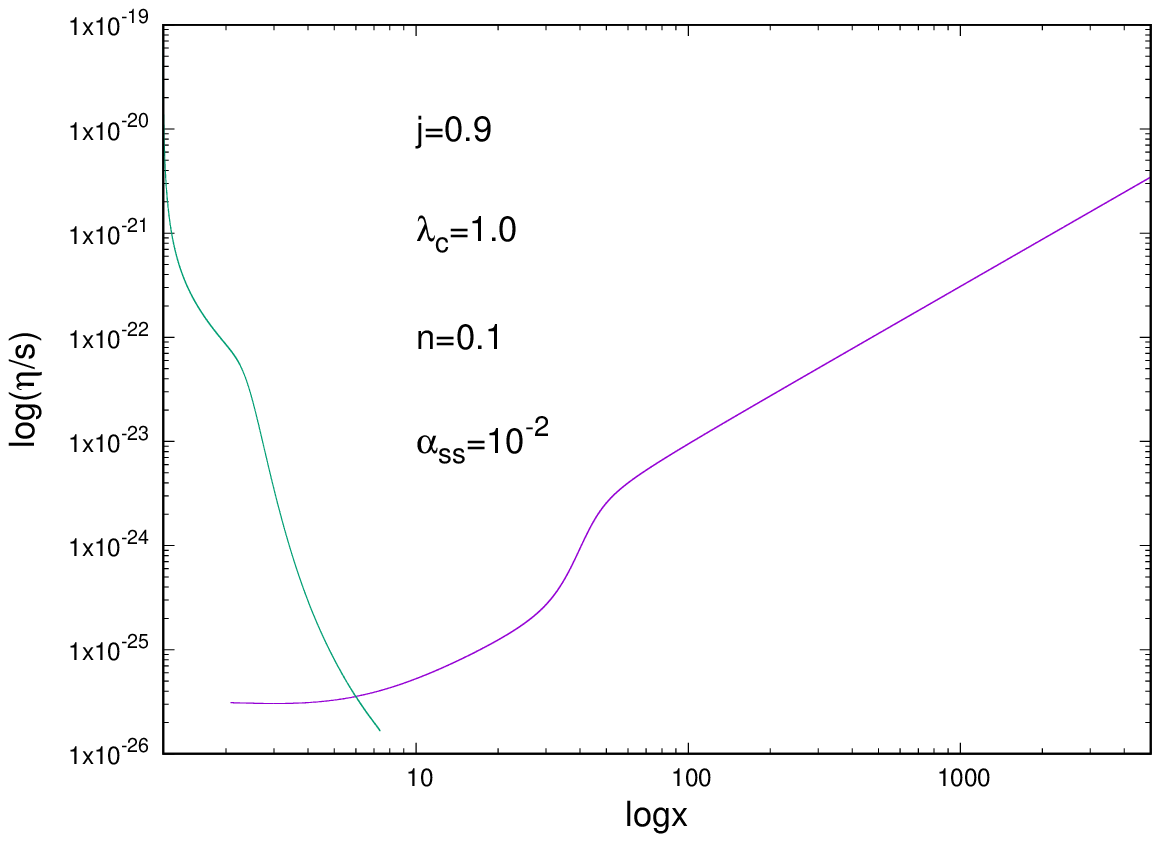}~~\\
	\it{Figures $3.1.a$ and $3.2.a$ are plots of $log(\frac{\eta}{s})$ vs $log(x)$ for viscous accretion disc flow with viscosity $\alpha_{ss}=10^{-4}$ and  $\alpha_{ss}=10^{-2}$ respectively, around a rotating BH with $j=0.9$ for adiabatic case. Figures $3.1.a$ and $3.2.a$ are plots of $log(\frac{\eta}{s})$ vs $log(x)$ for viscous accretion disc flow with viscosity $\alpha_{ss}=10^{-4}$ and  $\alpha_{ss}=10^{-2}$ respectively, around a non-rotating BH with $j=0.5$ for MCG. Accretion and wind curves are depicted by purple and green lines respectively.}
\end{figure}

In figures $1.2.a$ and $1.2.b$ we observe that this ratio is $10^2$ times higher as Shakura Sunyaev viscosity parameter $\alpha_{ss}$ is also increased by $10^2$ times (from $10^{-4}$ to $10^{-2}$). For these cases (point to be noted from equation (\ref{etaovers}), $\frac{\eta}{S} $ is proportional to $\alpha_{ss}$) while we increase the rotation of the central gravitating object, the lower limit of the ratio $\frac{\eta}{s}$ is obtained.
We can clearly distinguish between adiabatic flow and modified Chaplygin gas. Specially the wind branch completely changed its feature if we replace the accreting fluid by modified Chaplygin gas. The shear viscosity is proportional to the Shakura-Sunyaev parameter ($\alpha_{ss}$). So when viscosity is low the ratio $\frac{\eta}{s}$ becomes near to its lower bound for modified Chaplygin gas, but if the viscosity has increased then the ratio has failed to obtained its lower bound.

 Kovtun et al. \cite{kovtun2005} showed that the universal lower bound of the ratio $\frac{\eta}{s}$ is $\sim 6.08 \times 10^{-13}~~ Ks$ for a wide class of thermal quantum field theories. Mukhopadhyay \cite{mukhopadhyay2013} found that if the accretion flow is magnetically dominated or producing huge entropy due to the viscosity then the ratio $\frac{\eta}{s}$ is closer to its lower bound for the cases of hot, optically thin accretion flows. In this we have tried to study the ratio $\frac{\eta}{s}$ for viscous accretion disk around both rotating and non-rotating BHs in the presence of DE (here we take MCG as the DE representative). For adiabatic case the lower bound of the ratio $\frac{\eta}{s}$ is higher than the universal lower bound for viscous accretion flow. The scenario totally changed when we have introduced DE in the flow. The lower bound of the ratio $\frac{\eta}{s}$ surprisingly decrease beyond universal lower bound. In some recent articles \cite{Sandip1,Sandip2} authors have shown that DE catalyze the effect of viscosity in the accretion disk as DE exerts negative pressure. So from the equation \ref{entropy_density}), the value of  entropy density take very high value and the ratio $\frac{\eta}{s}$ becomes very very low.  
 
  In the background of reported absorpation signal (e.g. EDGES \cite{bowman_2018}), the dissipative effect of Dark Matter(DM), DE along with the DM-gas interaction is studied in several articles. In spite of considering ideal nature of exotic/ dark fluid, viscous nature has been considered which draws drastic changes in justification of the standard cosmological history. Decay in Cold Dark Matter (CDM) into relativistic particles is a procedure to give birth of bulk viscosity \cite{mathews_2008}. According to the \cite{velten_2013}, the value of shear viscosity is mainly constrained by the basic two properties of universe, viz., homogeneity and isotropy. This is why the shear viscosity for dark sector is found to be very small in value. Presently, a constain of dissipation of gravitational waves from $GW150914$ \footnote{Data obtained by the LIGO Hanford detector for the gravitational wave event on September $14$, $2015$ at $09:50:45$ UTC} is imposed, on the non zero value of shear viscosity \cite{goswami_2017}. The range of the shear viscosity found for $GW150914$, even does match with that of dissipative DM in galaxy clusters such as Abell $3827$ \cite{masseyet_2015}. In the path of $GW150914$, the cosmological fluid, let take it to be the DE, has upper bound of shear viscosity $\eta \lesssim 5.2 \eta_{crit} \sim 2.3 \times 10^9$ Pa Sec in $1 \sigma$CL, where $\eta_{crit}= \rho_{crit} H_0^{-1}= 3.21 \times 10^{-5} (GeV)^3 = 4.38\times 10^8$ Pa Sec, using $\rho_{crit}= 3 H_0^2 M_{PL}^2$. Clearly this value is very less. This can be understood if we imagine the extremely less self attraction between the DE components due to the negative pressure exerting nature of it. Whenever a fluid is expanding due to a self generated negative pressure, the attraction between its microstates will be lesser and the corresponding value of shear viscosity will decrease.
 
Now we will deviate our mind towards the entropy study. It should be noted that in this letter we are mainly assuming the phenomena of DE accretion on a supermassive black hole (SMBH). So what we need to consider is a global vast thing like DE along with its nature near to the region of a very local object called a SMBH which are generally located in the central region of a galaxy or galactic cluster etc. Initially, our universe comprised low entropy due to the low value of the initial gravitational entropy \cite{penrose_1989}. The standard cosmological structure has a cosmic event horizon and an associated Gibbon \& Hawking \cite{gibbons_1977} entropy as 
$$S_{CEH}(t_0)= \frac{kc^3}{G \hbar} \frac{\Delta}{4} = \frac{\pi k c^3}{G \hbar} R^2_{CEH} (t_0)^2~~,$$
where $$R_{CEH}(t)= a(t) \int_{t}^{\infty} \frac{c dt'}{a(t')}~~,$$
$a(t)$ being the scale factor. The entropy budget of different objects are listed below \cite{egan_2010} in Table $1$,
\begin{table}[h!]
	\begin{center}
		\caption{Entropy of different components of universe}
		\begin{tabular}{|c|c|}
			\hline
			\textbf{Component} & \textbf{Entropy }\\
			\hline
			Cosmic event horizon & $2.6 \pm 0.3 \times 10^{122}$\\
			SMBH & $1.2^{+1.1}_{-0.7} \times 10^{103}$\\
			Stellar BHs $(42 - 140 M_\odot)$ & $1.2 \times 10 ^ {98^{+0.8}_{-1.6}}$\\
			DM & $6\times 10 ^{86 \pm 1}$\\
			Stars & $3.5 \pm 1.7 \times 10^{78}$\\
			\hline
		\end{tabular}
	\end{center}
\end{table}
So DE, which is permeated all over inside the universe shows high entropy. Besides when we are to calculate the entropy density, as we are considering the comparatively small accretion zone, the value of entropy should be divided by small volumetric area which will keep the entropy density's volume high. Finally, we will come to derive the $\frac{\eta}{s}$ ratio which is a very small quantity, when will be divided by a large number. Whenever baryonic matter is considered (the highest entropy can be occured in the core of a star) the entropy is much less \cite{egan_2010} and the shear viscosity is high as baryonic matter much opposes the distortion in the volume. This is why we get the ratio of $\frac{\eta}{s}$ lower than the prepredicted lower limit of it.
 
 In a nutshell, we have calculated the shear viscosity to entropy density ratio for viscous accretion flow. We find that the lower bound of the ratio is $10^{-28}$ Ks, when exotic matter like modified Chaplygin gas is accreted towards a central gravitating object. The lower bound of the ratio which we obtained in this letter, is lower than the universal range. DE effects the four fundamental forces, i.e., gravitation, electromagnitism, strong force and weak force but the baryonic properties of DE are unknown to us. DE also reduces the attracting force between two baryonic/ accreting particle, even if we assume DE as group of particles then it creates repulsive force between two DE particles. As a result the frictional force in between the particles decreses which indicate that shear viscosity of DE accretion flow reduces. On the other hand our universe is expanding which means that entropy of our universe is also rapidly increasing. Since our universe is dominated by DE, the entropy density of DE accretion is very high that the ratio of shear viscosity to entropy density obtains its lowest value.

{\bf Achkowledgement : } This research is supported by the project grant of Goverment of West Bengal, Department of Higher Education, Science and Technology and Biotechnology (File no:- $ST/P/S\&T/16G-19/2017$). SD thanks to Department of Higher Education and technology, Goverment of West Bengal for research fellowship. RB thanks IUCAA for providing Visiting Associateship. RB also thanks Prof. Banibrata Mukhopadhyay, Department of Physics, IISc, Bangalore, India-560012 for detailed discussion regarding this problem earlier.

RB dedicates this article to his supervisor Prof. Subenoy Chakraborty, Department of Mathematics, Jadavpur University, Kolkata-32 on his $60^{th}$ birth year.

\end{document}